\documentclass[a4paper]{article}

\usepackage{INTERSPEECH2019}
\usepackage{amsmath,graphicx}
 \usepackage{epsfig,amssymb,multirow,boldline,graphicx,makecell,soul,hyperref,subfig,float}

\usepackage{xcolor}

\title{Large-scale Speaker Retrieval \\on Random Speaker Variability Subspace}
\name{Suwon Shon$^1$, Younggun Lee$^2$, Taesu Kim$^2$}
\address{Massachusetts Institute of Technology, Cambridge, MA, USA$^1$  \\
Neosapience, Inc., Seoul, South Korea$^2$}
\email{swshon@csail.mit.edu \{yg,taesu\}@neosapience.com}

\begin{document}

\maketitle
\begin{abstract}
This paper describes a fast speaker search system to retrieve segments of the same voice identity in the large-scale data. A recent study shows that Locality Sensitive Hashing (LSH) enables quick retrieval of a relevant voice in the large-scale data in conjunction with i-vector while maintaining accuracy. In this paper, we proposed Random Speaker-variability Subspace (RSS) projection to map a data into LSH based hash tables. We hypothesized that rather than projecting on completely random subspace without considering data, projecting on randomly generated speaker variability space would give more chance to put the same speaker representation into the same hash bins, so we can use less number of hash tables. Multiple RSS can be generated by randomly selecting a subset of speakers from a large speaker cohort. From the experimental result, the proposed approach shows 100 times and 7 times faster than the linear search and LSH, respectively.
\end{abstract}

\section{Introduction}

Today, so to speak, we are living in the flood of online videos and social media. It is known that hundreds of hours of video are uploaded every minute on Youtube.
Since most of the online videos include speech segments, collecting and abusing someone’s voice becomes much easier than ever before. Moreover, recent significant advances in speech synthesis technology made it easier to clone voice identity and it becomes harder to distinguish between original speech and synthesized one~\cite{Wang2017,Jia2018}.
In the area of multimedia retrieval, music and video search technology have been developed to protect the copyright of the contents. However, it is yet difficult to search someone’s voice identity from a large amount of multimedia data. Voice identity search would play an important role to prevent illegal cloning and abusing. One might apply conventional speaker identification or verification algorithms for such purpose. It could be done by training the speaker model and inferring the model score for every speech segment available online. But, it is practically not doable because of the large search space. Thus, it is inevitable to develop faster algorithms without performance degradation to prompt to a user as quickly as possible~\cite{Aronowitz2005}.


After audio segment can be represented in a single low-dimensional latent vector such as i-vector~\cite{Dehak2011}, hashing approach was proposed on the speaker search problem to maximize the speed of retrieval while minimizing the performance degradation~\cite{Jeon2012,Schmidt2014, Leary2014, Sturim2016}. Locality Sensitive Hashing (LSH)~\cite{Indyk1998} is a powerful tool to approximate the nearest neighbor search in high dimension and successfully adopted on top of the i-vector for a large-scale fast speaker identification.

For a large scale application, long hash bits are needed to reduce the size of each hash bin and to speed up a retrieval time. However, since the LSH is data-independent unsupervised hashing method rely on the random projection, it suffers from the redundancy of the hash bits. Also, long hash bits decrease the collision probability between similar samples. Consequently, we are encouraged to use a lot of multiple hash tables and this increases the query time and storage consumption. To reduce this redundancy, a supervised hashing method can be considered. A spectral hashing~\cite{Weiss2008} is a representative method in a supervised hashing method. It produces a very efficient hash code by learning a hash function from the training data. Ideally, all the true neighbor of an arbitrary query can be found in the specific hash bucket. However, this is not feasible using only a single hash table for a practical situation such as large-scale data lies on the high-dimensional space.

In this study, we propose a Random Speaker-variability Subspace (RSS) projection approach based on the Linear Discriminant Analysis (LDA) to give maximum efficiency on the hash table of LSH. Providing weak supervision by selecting random subset speakers in the training dataset when we generate random projection matrix for LSH, we observed it could dramatically save the hash bits (hyperplanes) and hash tables by removing the redundancy. We also re-define the speaker search task into more detail sub-task as speaker identification task and speaker retrieval task considering the query and search space. Experiments were conducted using both i-vector and neural network based speaker embedding, x-vector~\cite{DavidSnyder2017inter}. We used Voxceleb 1 and 2 dataset~\cite{Nagraniy2017, Chung2018} and prepared a various combination of dataset categories to measure the performance of speaker search sub-tasks.


\label{sec:intro}

\section{Speaker search in the wild}
Speaker search can be divided into two sub-categories as speaker identification and speaker retrieval. The main difference between two sub-tasks is whether the query is from arbitrary speech segment or target speaker segment.
\label{sec:spkid}
\subsection{Speaker Identification}
For speaker identification, the query $x$ is a latent vector from an arbitrary speech segment and its class (speaker index) is $C_x$. Search space is $D=\{\omega_1,\omega_2,...\omega_S\}$ where $S$ is total number of Speaker and $\omega_s$ is a latent vector from a speaker speech segment, $s$. $y_s$ is the score between $x$ and $\omega_s$. 
For closed-set speaker identification which query speech is assumed to be spoken by one of the speakers in the search space, the system forced the input query to match the best matching speaker in the search space as $C_x = C_{y^*}$ where $y^* = \underset{s={1...S}}{\mathrm{max}} y_s$. Previous study on large-scale speaker identification mostly supposed this closed-set identification to measure the performance~\cite{Jeon2012,Schmidt2014}. 

However, in practical situation, the system needs to consider that the speech query may not be spoken by the speaker in the search space. This can be represented as open-set speaker identification. Open-set speaker identification is also known as multi-target detection, so we can borrow the top-1 stack detector performance measurement term in~\cite{Singer2004} about miss and false alarm probability.






\subsection{Speaker Retrieval}
\label{sec:speaker_retrieval}

For speaker retrieval, the query $\omega$ is a latent variable from a speech segment of the target speaker we are interest in and its class is $C_\omega$.
The search space is $D=\{x_1,x_2,...x_N\}$ where $N$ is the total number of segments to search and $x_n$ is a latent vector from an arbitrary speech segment from an unknown speaker. 
Compared to the speaker identification task, the query and search space are changed each other. The aim of this task is to retrieve all relevant documents. Thus, speaker retrieval is 1 to N verification task. The performance measurement is the same as the 1 to 1 verification task which is a general speaker verification task. However, the trials to measure the performance need to be prepared by full combination between the query and search space. For instance, if we have 10 target speakers for an input query, the number of trials is 10$\times N$.






\section{Speaker Dataset}
\label{sec:data}
To make up speaker search environment, we used Voxceleb 1 and 2 dataset~\cite{Nagraniy2017, Chung2018}. The Voxceleb dataset is composed of automatically collected audio and video data for large scale speaker identification. The collecting pipeline includes face detection and active speaker verification to verify all video clips have detected faces with synchronized speaking voice by the conservative threshold to maximize the precision. These processes are reasonable pre-processing to our speaker search scenario, so we used the audio part of the Voxceleb dataset without modification. Voxceleb 1 and 2 have more than 1,281,352 utterances from 7,365 speakers. Utterances were extracted from a video clip and each clip has roughly 10 to 50 utterances. We used Voxceleb 1 development set (147,935 utterances from 1,211 speakers) for training on both identification and retrieval task. For the test, we combined the utterances in a different way for each task.

\noindent
\textbf{Speaker identification task}: For search space in the speaker identification task, we used the utterances from first 3 video clips in Voxceleb 1 test set, plus entire utterances from Voxceleb 2 development set. We will use all utterances from the same speaker at once for speaker representation, so each entity of search space represent a unique speaker and include total 6,034 speakers.
For the query, we used the rest utterances in the Voxceleb1 test set. Since we already picked the utterances from first 3 clips for search space, all the query must have the same identity in search space and this means closed-set identification task as previous studies~\cite{Jeon2012,Schmidt2014}.

\noindent
\textbf{Speaker retrieval task}: For search space, we used 100,000 utterances from Voxceleb 2 development set and 3,776 utterances from the first 3 clips in Voxceleb 1 test set. We randomly selected 100,000 utterances in the Voxceleb 2 development set from 1,092,009 utterances, so total 103,776 utterances became the search space. For the query, we used utterances from first 3 clips in Voxceleb 1 test set (total 40 speakers, a target class which has relevant utterance in search space) and Voxceleb 2 test set (total 120 speakers, a non-target class which does not have relevant utterances in search space). Thus, 160 speaker representations were used as query.

We mainly focused on the speaker retrieval task, but we also conducted an experiment on speaker identification task as the previous study reported~\cite{Schmidt2014}.

\begin{table}[]
\centering
\caption{Speaker verification performance by speaker representation method on Voxceleb1 test set}
\label{tab:baseline_sv}
\resizebox{0.18\textwidth}{!}{%
\begin{tabular}{c|c}
\hlineB{2}
Speaker & \multirow{2}{*}{EER} \\
representation &  \\ \hlineB{2}
i-vector (cosine) & 8.1 \\ 
i-vector (PLDA) & 5.4 \\ 
x-vector (cosine) & 9.9 \\ 
x-vector (PLDA) & 6.0 \\ \hlineB{2}
\end{tabular}%
}
\end{table}

\section{Fast large-scale speaker retrieval}

\subsection{Speaker recognition system}
For performance comparison, we considered two speaker representation, i-vector~\cite{Dehak2011} and x-vector~\cite{DavidSnyder2017inter}. To train i-vector and x-vector, we used the voxceleb 1 development set as depicted in the section~\ref{sec:data}. We followed other training detail same as~\cite{Shon2018frame}. Table~\ref{tab:baseline_sv} shows the speaker verification performance of i-vector and x-vector using Equal Error Rate (EER) measurement. We used the same system as reported in~\cite{Shon2018frame}. Note that since the Voxceleb 1 contains only 1,211 speakers which is a small number to training DNN, x-vector shows slightly worse performance than i-vector. 

\subsection{LSH}
The goal of our study is that to retrieve similar speaker efficiently with minimum performance loss. In the previous study, LSH was adopted for this task and showed very promising performance on the large-scale speaker identification task combining with i-vector. 

LSH projects the i-vector into random hyperplane. This hash operation maps close vectors into the same bins(i.e. buckets) with high probability~\cite{Gionis1999}. Suppose $r$ is $d$-dimension random projection vector drawn from a standard normal distribution, $d$ is dimension of original i-vector $w$. The hash function maps i-vector $w$ as :
\begin{equation}
    w_{r}=h_r(w)=\text{sgn}(w^Tr)=
    \begin{cases}
    1 & \quad \text{if } w^Tr\geq0 \\
    0 & \quad \text{if } w^Tr<0
  \end{cases}
\end{equation}
We can concatenate several hash functions and use multiple, independent hash functions to boost performance, by using $d\times k$ dimensional random projection matrix $R_l$ where $k$ is the number of hyperplane per hash table, $d$ is dimension of original speaker representation, $l$, $1\leq l\leq L$, is the hash tables index and $L$ is the total number of hash tables. Thus, the parameter $k$ and $L$ need to be chosen carefully since the parameter decides the trade-off between performance and computational efficiency.

\begin{figure}[ht]
    \centering
    \subfloat[Random projection]{\includegraphics[width=0.4\linewidth]{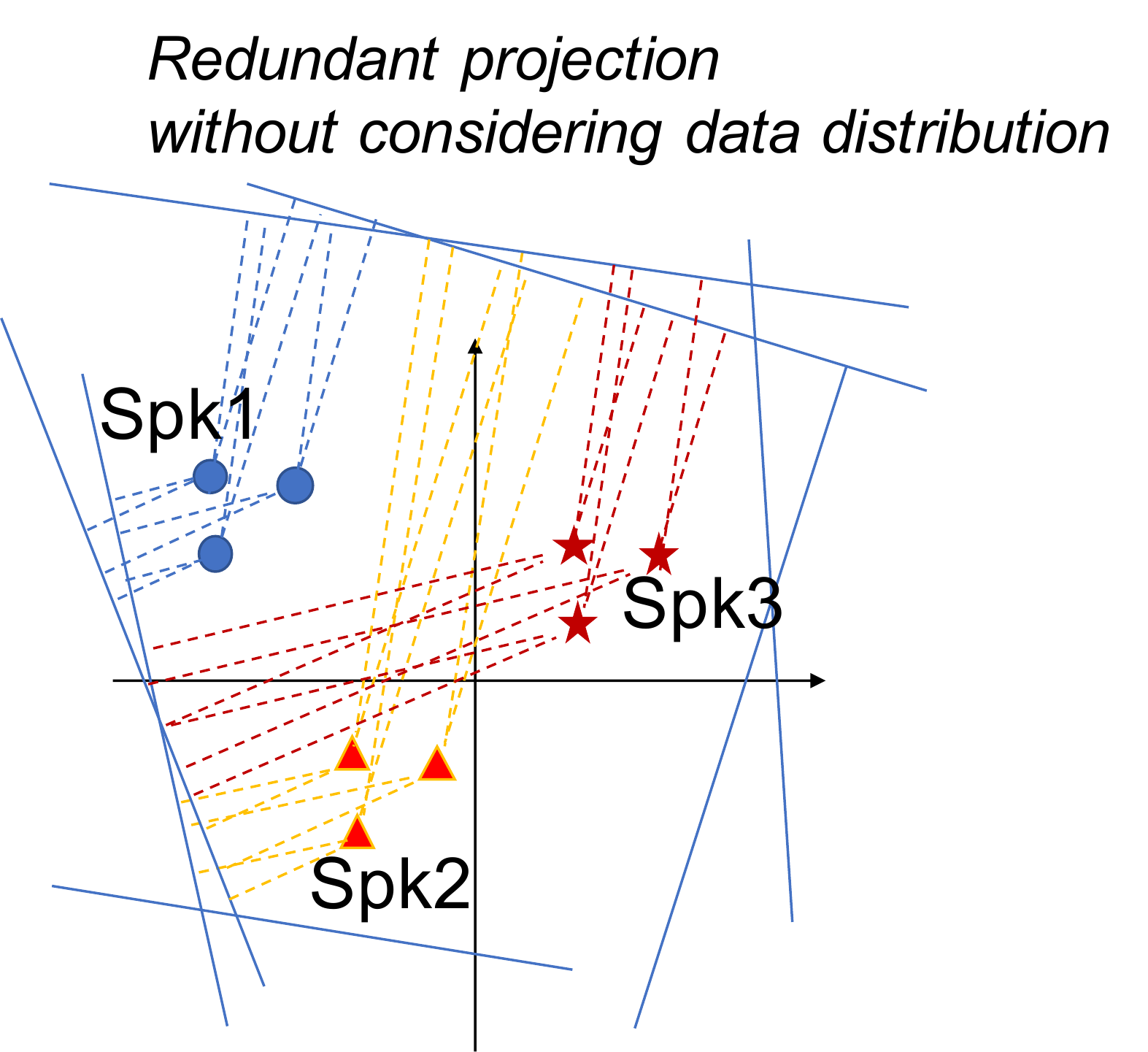}}
    \hspace{1cm}
    \subfloat[RSS projection (No. of random speaker=2)]{\includegraphics[width=0.4\linewidth]{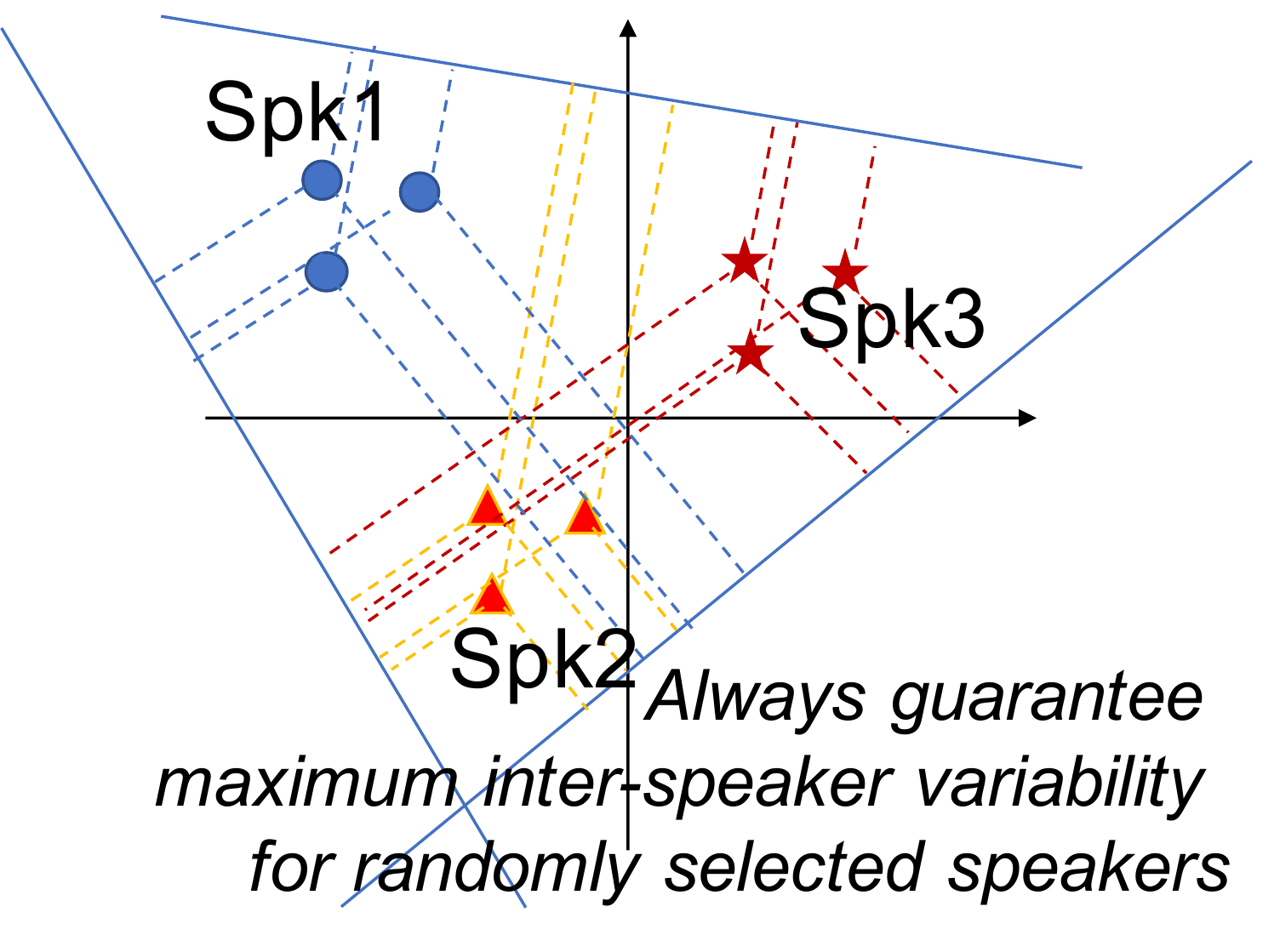}}
    \caption{Example of projections for 3 speakers. Same color represents different utterances for same speaker.}
    \label{fig:projection_examples}
\end{figure}
In this study, we basically follow the same i-vector retrieval with LSH algorithm in~\cite{Schmidt2014} and also used cosine distance which guarantees its performance in conjunction with LSH. We measured the baseline performance of closed-set speaker identification and speaker retrieval tasks using i-vector and x-vector as shown in table~\ref{tab:baseline_sr}.

\begin{table}[]
\centering
\caption{Speaker search performance using i-vector and x-vector}
\resizebox{0.4\textwidth}{!}{%
\begin{tabular}{c|c|c}
\hlineB{2}
Speaker & Speaker & Speaker \\
Representation & retrieval, EER(\%) & identification, Acc.(\%) \\ \hlineB{2}
i-vector & 4.36  & 79.23 \\ 
x-vector & 5.60  & 76.74 \\ \hlineB{2}
\end{tabular}%
}
\label{tab:baseline_sr}
\end{table}

\begin{figure}[ht]
\centering
\includegraphics[width=0.77\linewidth]{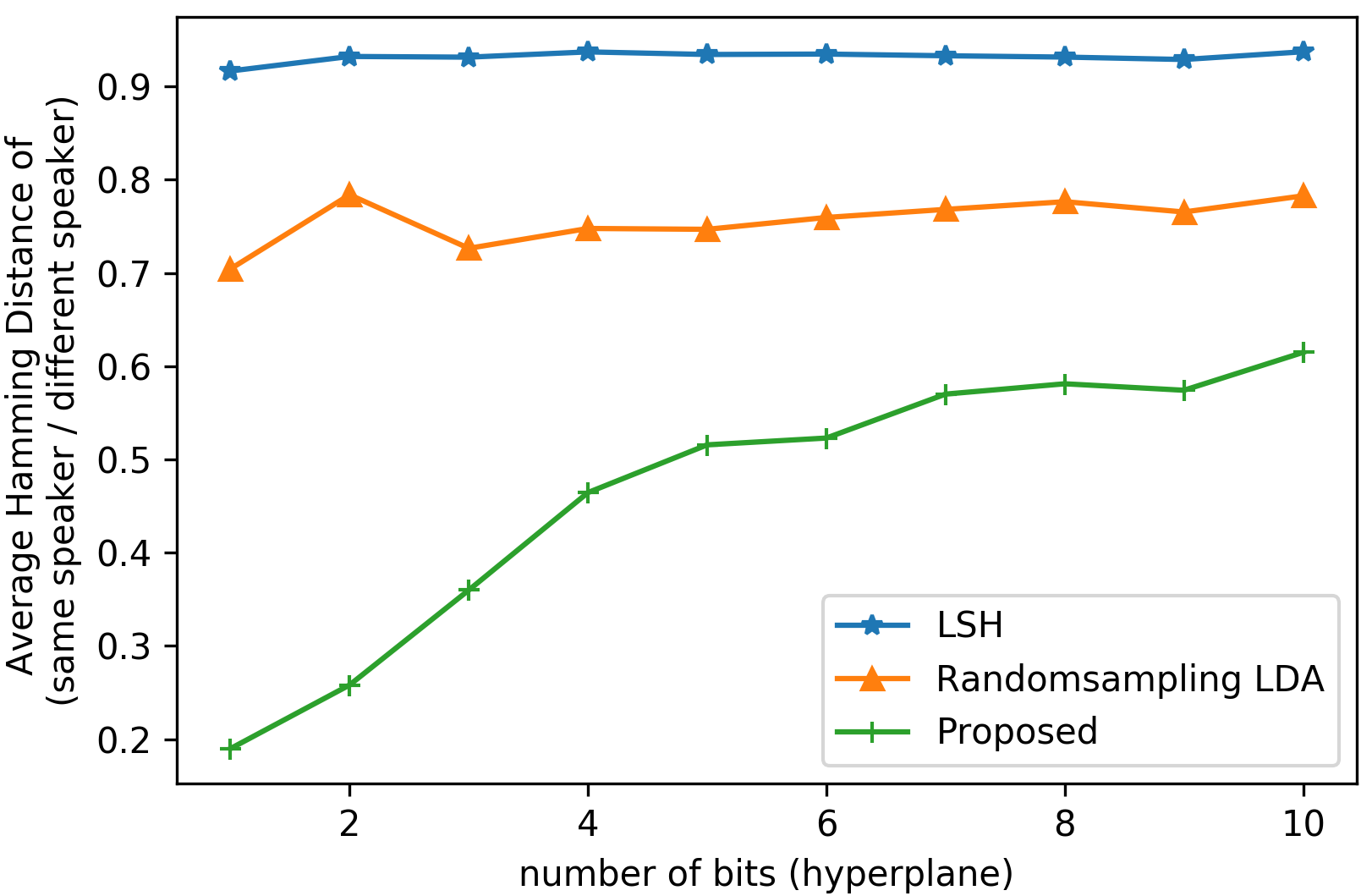}
\caption{Average Hamming distance of same identities divided by the distance of different identities. Lower is better.}
\label{fig:hamming}
\end{figure}


\begin{figure*}[ht]
    \centering
    \subfloat[Varying number of hyperplane, $k$]{\includegraphics[width=0.4\linewidth]{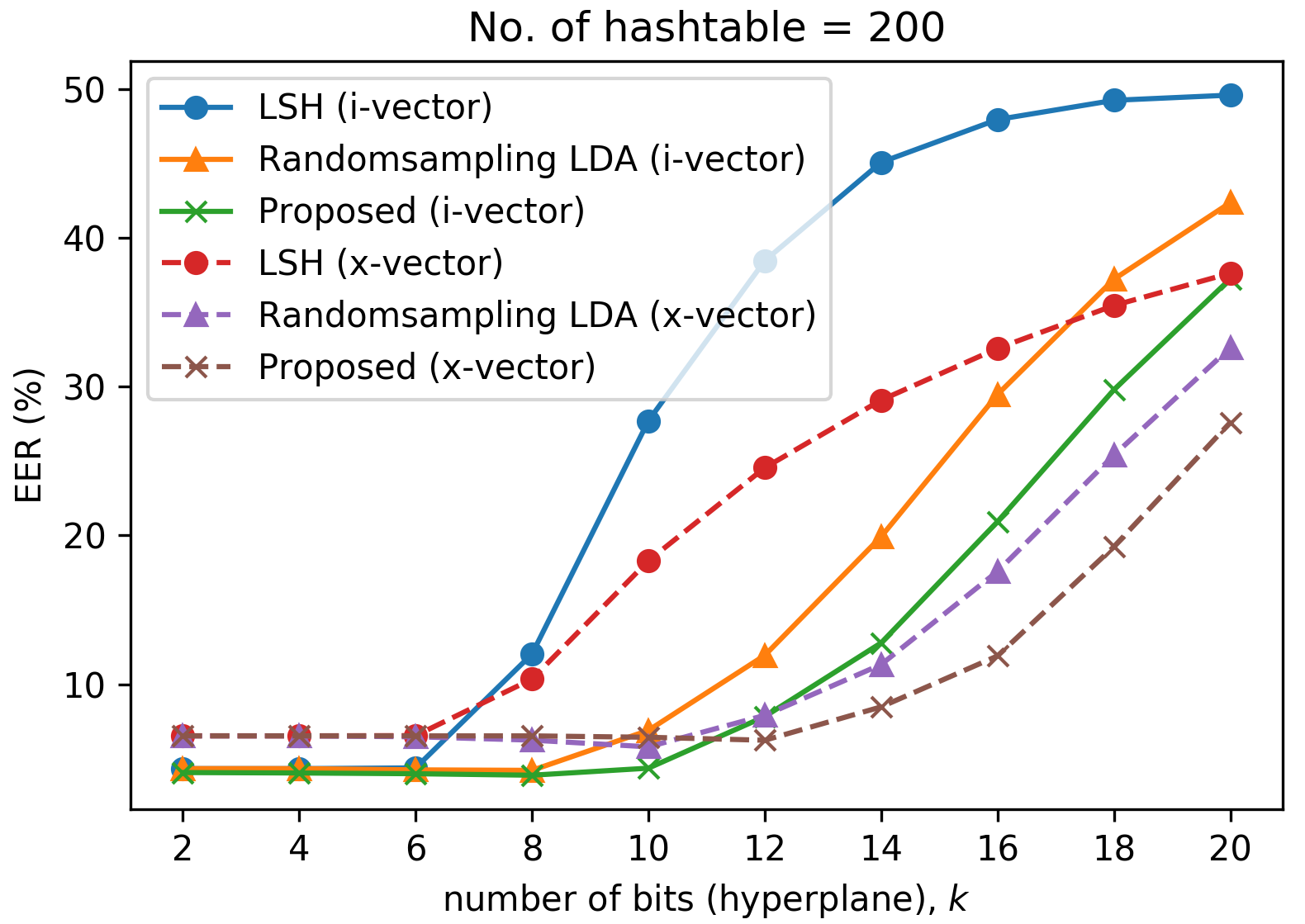}}
    \hspace{1cm}
    \subfloat[Varying number of hash table, $L$]{\includegraphics[width=0.4\linewidth]{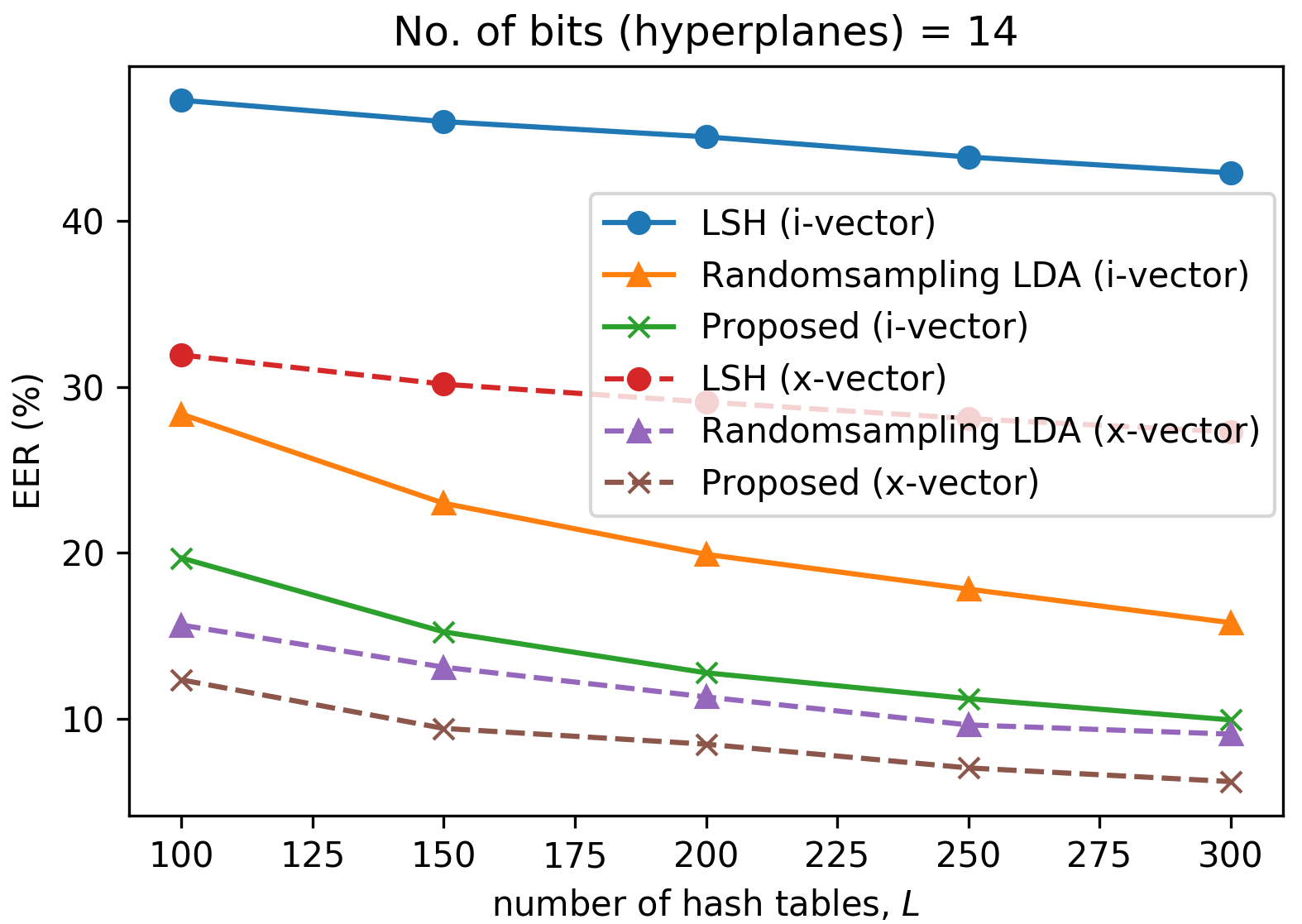}}
    \caption{EER measurement on speaker retrieval task.}
    \label{fig:fig_perform_byparamter}
\end{figure*}

\subsection{Random Speaker-variability Subspace (RSS) Projection}

Since LSH is an unsupervised hashing method that uses random projection matrix drawn from a $d$-dimensional standard normal distribution, it suffers from redundancy of the hyperplanes (hash bits) as represented in figure~\ref{fig:projection_examples}. Thus, lots of hash tables are needed to access enough points for the satisfactory recall. In this situation, we can give a weak supervision to generate the random subspace projection matrix using speaker labels to make the same speaker's voice mapped into the same bin more efficiently. We generate Linear Discriminant Analysis (LDA) transformation matrix directly by using utterances from the random subset of the speakers and use this matrix for projection matrix $R_l$.

Suppose $S_l$ is a randomly selected speaker subset from training dataset where the number of speaker is $N_s$. Using the speaker representation in the subset $S_l$, we can obtain the between-class scatter matrix and the within-class scatter matrix. Then, we can calculate LDA matrix which maximizes the ratio of the two scatter matrices. This LDA matrix project speaker representation into RSS and we can generate it $L$ times randomly to substitute the random projection matrix $R_l$ in LSH. For hyperparameter $N_s$, if we choose too many speakers, it will generate a similar LDA matrix which have many redundancy between projection matrices. Also we have to choose $N_s$ more than the length of hash bits $k$ to project into $k$ dimension. To balance the size of each hash bin, we modified the hash function as $h_r(w)=\text{sgn}(w^Tr+b)$ where $b$ is the mean of projected data, i.e, $b=-\tfrac{1}{N}\sum_{i=1}^N{w_i^Tr}$


Similar work was reported in the face recognition study~\cite{Wang2004} that uses random sampling LDA.
They generated multiple projection matrix by selecting eigenvectors randomly. Then LDA was applied on each projected subspace.
Since they have only $d$ eigenvectors to be selected randomly, the redundancy between the projected subspaces would be increased if we increase the number of hash tables.
This approach is not actually designed for hashing. However, projecting samples onto random subspace to enforce weak classifiers is similar scheme to our approach, so we also use this random sampling LDA to generate projection matrix $R_l$ and compare it with other approaches.

In the projected subspace, hamming distance approximates the cosine distance as the number of random hyperplanes $k$ is increases~\cite{Jansen2012}:
\begin{equation}
\cos{(w_i,w_j)} \approx \cos{(\frac{H(h_r(w_i),h_r(w_j))}{k}\pi)}
\end{equation}
where $H(\cdot)$ is a hamming distance. Thus, if the projection matrix effectively approximates the original distance, the speaker representation from the same identity has high possibility to be in the same or near bucket and we can use less number of hash functions to approximate the original distance. It means that the hamming distance of speaker representation from the same identity would be closer than others. Taking this into consideration, the approximation capability of a new projection matrix can be measured by the averaged hamming distance of the same identities and different identities. We checked this averaged hamming distance using LSH, random sampling LDA, and the proposed method. As shown in figure~\ref{fig:hamming}, the proposed method shows that the distance between the same speakers is close, but the different speakers result in far distance. From this observation we can expect the proposed method would give significant efficiency than others.

\section{Experimental results}

We conducted two tasks of experiments, speaker identification and speaker retrieval, and the experimental environments were fully described in the section~\ref{sec:data}. For speaker retrieval, EER was used for performance measurement, for the speaker identification task (closed-set), accuracy was used for as described in section~\ref{sec:spkid}. I-vector and x-vector were extracted in 600 and 512 dimensions respectively, then reduced into 150 dimensions using LDA.
We measured the retrieval time to return candidates for a given query and excluded the time to extract i-vector and x-vector.
For the baseline, we generate $L$ random projection matrix which drawn from $d$-dimensional standard normal distribution.
For RSS projection, the optimal operating $N_s$ is varied by the number of hyperplanes. Rather than optimizing, we set to $d$, equals to i-vector or x-vector dimension, so the subset has more than $d$ samples at least to avoid within-class scatter matrix become singular by small sample size problem~\cite{Huang2002}. 
We repeated $L$ times to generate RSS projection matrix by selecting random speakers for each matrix. 
For the random sampling LDA~\cite{Wang2004}, we used randomly 100 eigenvectors from PCA for random subspace, then repeat $L$ times to generate random sampling LDA matrix. 
We used these three projection matrix $R_l$ for LSH.

\begin{figure}[ht]
    \centering
    \subfloat[Speaker retrieval task. Y-axis represent EER(\%) value.]{\includegraphics[width=0.75\linewidth]{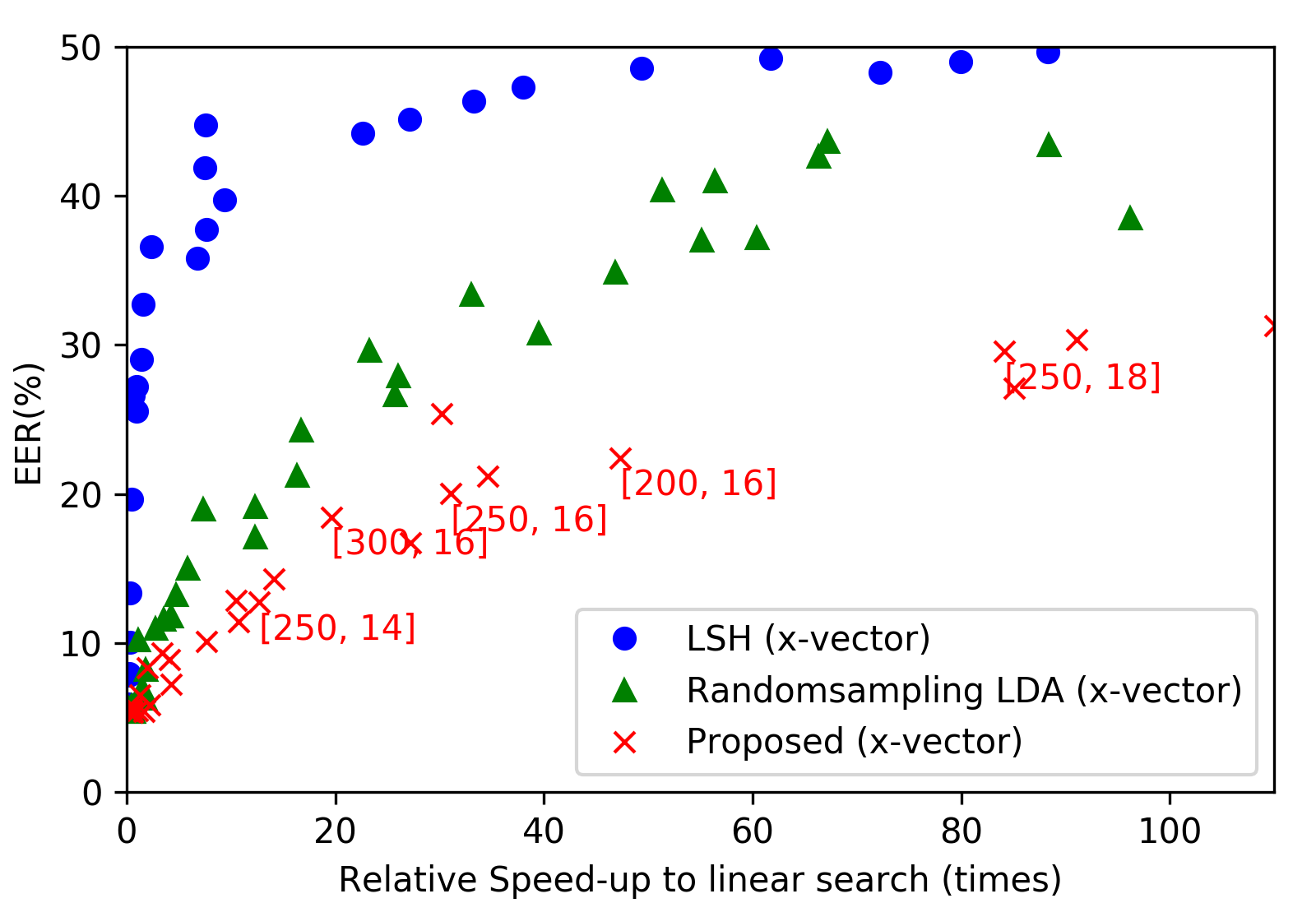}}
    
    \subfloat[Speaker identification task. Y-axis represent accuracy relative to linear search accuracy.]{\includegraphics[width=0.78\linewidth]{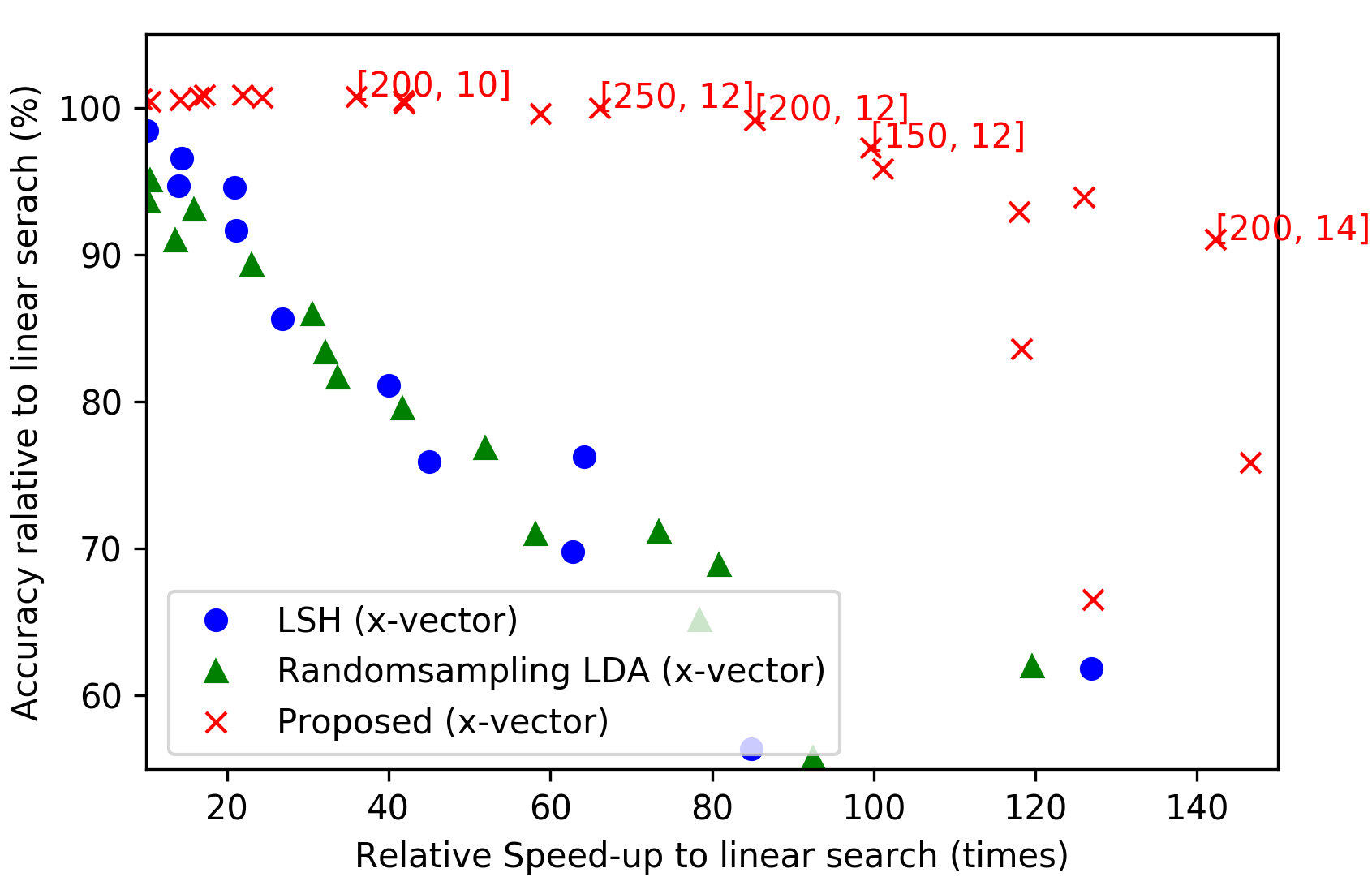}}
    \caption{Speedup vs. performance trade-off. The number of $k$ varied from 2 to 20 in step of 2 and $L$ from 100 to 300 in steps of 50. A pair of numbers on the plot represent $[L,k]$ as examples.
    }
    \label{fig:fig_perform_tradeoff}
\end{figure}

Figure~\ref{fig:fig_perform_byparamter} (a) and (b) shows the result on various number of hyperplanes and hash tables. The proposed RSS projection shows significant efficiency compared to others in the EER measurement. Meanwhile, it is interesting that the x-vectors mapped into hash bins very efficiently than i-vectors in all methods. We surmise that this phenomenon is because the x-vector extracting DNN was trained discriminatively with one hot speaker label while the i-vector framework assumes that the i-vector distributed in Gaussian distribution. Thus, the distance between the same speaker x-vectors is more likely to become close than the i-vector. We believed that any speaker embeddings extracted from DNN would take this advantage, not only x-vector. Figure~\ref{fig:fig_perform_tradeoff} (a) and (b) shows trade-off plot between retrieval speed and performance. We conducted the experiments by varying the parameter $k$ and $L$ and scattered in speed and performance axis. On both speaker retrieval and identification task, the proposed approach shows remarkable performance improvement. 
For example, to speed up while maintaining above 95\% speaker identification performance relative to linear search, the proposed approach has 100 times faster than linear search and 7 times faster than LSH as shown in table~\ref{tab:final_summary}.
Note that the EER in figure~\ref{fig:fig_perform_byparamter} and figure~\ref{fig:fig_perform_tradeoff} (a) is absolute value, not relative value to linear search.
To give an intuition on the parameters, we specified pairs of parameters on the plot in [$L,k$] format by varying a parameter while the other was fixed. As shown in the figure, $k$ is more sensitive to performance than $L$. Thus, $L$ can be used for fine-tune to satisfy required performance.

\begin{table}[ht]
\centering
\caption{Performance summary of hashing methods on speaker identification}
\label{tab:final_summary}
\resizebox{0.4\textwidth}{!}{%
\begin{tabular}{c|c|c|c}
\hlineB{2}
Hashing & LSH & \begin{tabular}[c]{@{}c@{}}Random-sampling \\ LDA\end{tabular} & \begin{tabular}[c]{@{}l@{}}Proposed\end{tabular} \\ \hlineB{2}
Baseline Accuracy & \multicolumn{3}{c}{ 76.74 \%}   \\ \hline
Hashing Accuracy & 74.10\% & 74.14\% & 74.65\%  \\ \hline
Relative speed & 14$\times$ & 7$\times$ & 100$\times$ \\ \hline
No. of Hyperplanes & 10 & 6 & 12  \\ \hline
No. of Hash tables & 300 & 150 & 150  \\  \hlineB{2}
\end{tabular}%
}
\end{table}

\section{Conclusion}
We have proposed a RSS based hash algorithm to search and retrieve someone's voice identity, which is applicable to a large scale speech dataset. Previous studies have focused on the speaker identification, but we have redefined the search problem both in the speaker identification and retrieval. 
The proposed approach has shown significant efficiency to save retrieval time and storage consumption compared to ``vanilla" LSH that uses random projection. We have also observed that the speaker embedding is more advantageous to be mapped into a hash table compared to traditional i-vector.
To the best of our knowledge, this is the first study to use the class supervision on hashing for acoustic information retrieval. 
For future work, we would explore speaker retrieval methods further in many other domains as well as real-world applications. It would be also interesting to investigate how the system works in the presence of cloned voice by recent speech synthesis algorithms. 

\clearpage
\newpage
\bibliographystyle{IEEEtran}
\bibliography{template}

\end{document}